\documentclass[aps,reprint,onecolumn]{revtex4-2}
\usepackage[english]{babel}
\usepackage[utf8]{inputenc}
\usepackage{xcolor} 
\usepackage{physics}
\usepackage[colorinlistoftodos, color=green!40, prependcaption]{todonotes}
\usepackage{amsthm,amsmath,amssymb}
\usepackage{mathtools}
\usepackage{physics}
\usepackage{graphicx}
\usepackage[left=15mm,right=15mm,top=20mm,columnsep=15pt]{geometry} 
\usepackage{adjustbox}
\usepackage{placeins}
\usepackage[T1]{fontenc}
\usepackage{lipsum}
\usepackage{csquotes}
\usepackage{float}
\usepackage{soul}
\usepackage{subcaption} 
\usepackage{tikz}
\usepackage{comment}
\usepackage[pdftex, pdftitle={Article}, pdfauthor={Author}]{hyperref} 

\newcommand{\correctref}[1]{\textcolor{black}{#1}}

\begin{document}

\title{Resonance of an object floating within a surface wavefield}

\author{Sébastien Kuchly$^{1,\dagger}$, Wilson Reino$^{2,3\dagger}$, Kyle McKee$^{4}$, Stéphane Perrard$^{1}$, Giuseppe Pucci$^{3,5}$,Antonin Eddi$^{1}$}

  \email[Correspondence email address: ]{antonin.eddi@espci.fr \\ $^\dagger$Co-first author.}

  \affiliation{$^{1}$PMMH, CNRS, ESPCI Paris, Université PSL, Sorbonne Université, Université de Paris Cité, F-75005, Paris, France.}
  
 \affiliation{$^{2}$Dipartimento di Fisica, Università della Calabria, Via P. Bucci 31C, 87036 Rende (CS), Italy.}
  
  \affiliation{$^{3}$Consiglio Nazionale delle Ricerche - Istituto di Nanotecnologia (CNR-NANOTEC), Via P. Bucci 33C, 87036 Rende (CS), Italy.}
  
  \affiliation{$^{4}$Department of Mathematics, Massachusetts Institute of Technology, 02139, Cambridge, MA, USA.}

  \affiliation{$^{5}$INFN, Sezione di Lecce, Via per Monteroni, Lecce 73100, Italy.}

\date{\today} 

\begin{abstract}

We examine the interaction between floating cylindrical objects and surface waves in the gravity regime. Since the impact of resonance phenomena associated with floating bodies, particularly at laboratory scales, remains underexplored, we focus on the influence of the floats' resonance frequency on wave emission. First, we study the response of floating rigid cylinders to external mechanical perturbations. \correctref{Using an optical reconstruction technique to measure surface wave fields in both space and time, we study the natural resonance frequency of floats with different sizes. The results indicate that the resonance frequency is influenced by the interplay between the cylinder geometry and the solid-to-fluid density ratio. Second, these floating objects are placed in an incoming wave field. These experiments demonstrate that floats diffract incoming waves, while radiating secondary waves that interfere with the incident wavefield. Minimal wave generation is observed at resonance frequencies. These findings can provide insights for elucidating the behavior of larger structures, such as sea ice floes, in natural wave fields.}

\end{abstract}

\maketitle
\section{INTRODUCTION}


\correctref{ Wave-structure interactions are encountered in various engineering applications, such as naval architecture or offshore operations, including ships, platforms and buoys. \cite{falnes_ocean_2002,newman_marine_1977}. An external wave field may induce motion of a floating body possessing up to six different degrees of freedom. Linear translation modes in the horizontal plane are called sway and surge, while vertical translation is referred to as heave. Angular translation modes are called roll, pitch and yaw \cite{falnes_ocean_2002}. A number of theoretical studies have aimed to better understand these motions and the corresponding pressure forces on floating structures. Ursell \cite{ursell_heaving_1949} investigated the vertical heave motion of an infinitely long circular cylinder placed horizontally at the water surface, and computed the 2D velocity potential describing cylinder oscillations in the case of a non viscous purely potential flow. This enabled the computation of the added mass – the effective inertia added to the cylinder due to fluid which is displaced with the object's movement. Havelock \cite{havelock_waves_1955} generalized these results in three dimensions by looking at the heaving motion of a floating half-immersed sphere. The added mass was computed as a function of the oscillation frequency. Havelock also computed the evolution of the damping coefficient with the motion frequency. Forces exerted by an incoming, monochromatic wave field on a fixed floating body were then studied by Newman \cite{newman_exciting_1962}. Pressure forces have been computed for floating objects of various shapes including, for example, ellipsoids. Black \cite{black_radiation_1971} and Garett \cite{garrett_wave_1971} were interested by the forces exerted on a vertical circular cylinder excited by a monochromatic wave. Yeung \cite{yeung_added_1981} then computed the added mass and damping coefficient for the heave, surge and pitching motion of the same object. All these theoretical studies brought precious knowledge to better understand the behaviour of floating structures in small amplitude wave fields.}

~\\
\correctref{ The need for modelling polar climates and specifically wave interactions with sea ice motivated several studies on wave transmission by an array of floes. Indeed, sea ice is a brittle material that can be easily broken into fragments by surface waves, creating large areas of ice floes called Marginal Ice Zones \cite{squire_ocean_2020}. These collections of ice floes are known to effectively attenuate incoming waves \cite{squire_direct_1980}. One of the major mechanisms that explains this decay is the wave scattering by individual ice floes which was studied theoretically by Wadhams \cite{wadhams_attenuation_1988}. Each ice floe is put into motion by the incoming wave, generating radiative wave fields which can interact and result in the general damping of the initial wave field. These models and observations motivated several experiments performed in wave tanks such as the works of Bennetts and Williams \cite{bennetts_water_2015} or Yiew et al. \cite{yiew_hydrodynamic_2016}. Single floating discs were excited by a linear monochromatic wave. The disc motion was recorded using markers placed on the objects or accelerometers \cite{bennetts_water_2015}. This enabled the tracking of the amplitudes of heave, surge and pitch motion as a function of the wave frequency and amplitude. }

~\\
\correctref{ Recent interest for wave-energy converters also motivated several experimental studies. For instance, Kramer et al. \cite{kramer_highly_2021} and Colling et al. \cite{colling_free-decay_2022} studied the heave motion of a spherical buoy during a free-decay drop test. The buoy was dropped from a small height so that the water surface disturbance remained linear. The spherical buoy then underwent damped harmonic oscillations. Colling et al. observed that the oscillation frequency as well as the attenuation rate both depend on the buoy draft. Using a multi camera system as well as wave gauges they were able to compare the motion of the buoy and the local water surface deformation to different numerical models. While these large scale experiments mostly focus on the oscillation of the floating body and only study the water deformation at some point of the surface, experiments at a smaller centimetric scale enable the reconstruction of water surface around an oscillating body in both space and time. Rhee et al \cite{rhee_surferbot_2022} studied the waves generated by a centimeter-scale, vibrating floating robot thanks to Fast Checkerboard Demodulation (FCD) \cite{wildeman_real-time_2018}. This method is a variation of the Synthetic Schlieren Imaging method as developed by Moisy et al. \cite{moisy_synthetic_2009}, now widely used for wave measurement at the laboratory scale \cite{Chantelot2020, Domino_2020, doudic_2024}. Based on the difference of optical path length between a background pattern and the camera, these methods enable the computation of water surface distorsion. }

~\\
\correctref{ In this paper we focus our attention on the wave field generated by the motion of a floating circular cylinder. Using a surface reconstruction technique, we resolve in space and time the wave field radiated by a circular centimetric disk when submitted to an impulsive perturbation. From this wave field, we measure the natural resonance frequency of the object. Using the same reconstruction method, we place the floe in an anisotropic monochromatic wave and record the new wave field which results from the combination of the incoming, radiated and diffracted fields. Close to the floater's natural resonance frequency, we observe a minimal re-emission of waves in the direction transverse to the initial wave field. The manuscript is organized as follows. In Section II, we experimentally measure the resonance frequency of floating cylindrical bodies and propose a theoretical model. In Section III, we study the interaction of this bodies with an externally generated wave field, highlighting how the float's natural frequency affects this external field. }

~\\

\section{NATURAL FREQUENCY OF OSCILLATION OF A FLOATING OBJECT} 

\subsection{Experiments}\label{sec:experiments}

\begin{figure}[h!]
    \centering
    \includegraphics[width=0.80\columnwidth]{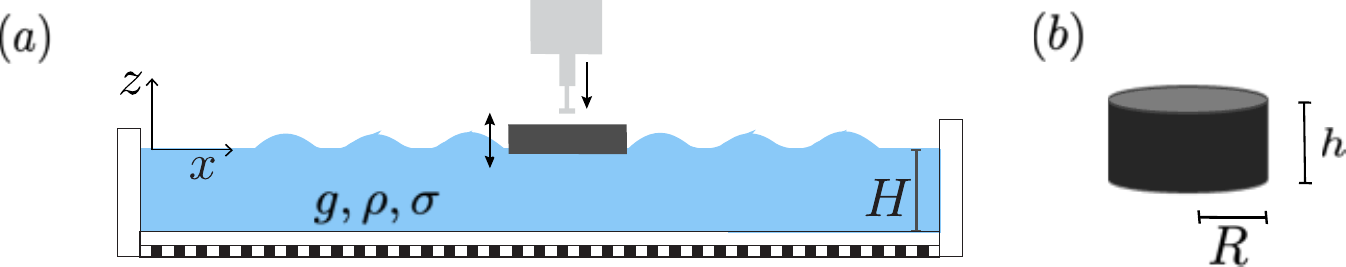}
    \caption{Experimental setup for measuring the natural frequency $f_0$ of a floating body. (a) Side-view schematic (not to scale) and (b) body geometry.}
    \label{fig:Schematic_resonance_frequency_setup}
\end{figure}
As floating objects, we use polyethylene cylinders with density $\rho = 0.95 \,\rm{g/cm^3}$. The cylinder radius $R$ ranges from $R = 7.5 \,\rm{mm}$ to $30.0 \,\rm{mm}$, and the height $h$ ranges from $h = 5.0 \,\rm{mm}$ to $20.0 \,\rm{mm}$ (Fig \ref{fig:Schematic_resonance_frequency_setup}b). Cylinders float at the surface of water in a tank with dimensions $78 \,\rm{cm} \times 38 \,\rm{cm} \times 19 \,\rm{cm}$. The water depth $H = 5.0 \,\rm{cm}$ ensures that the surface waves are in the deep water regime in the entire working frequency range. We investigate a relatively wide range of aspect ratios ($h/R$), from $0.17$ to $2.67$. Due to their thickness and Young's modulus, the floaters can be considered as rigid bodies in the working frequency range. To ensure reproducible wetting conditions and to prevent overwashing, we cover the external surface of each cylinder with adhesive tape. 

We first investigate the vertical oscillation mode (heaving) as a function of the floater dimensions. Starting from a floater at rest, we perturb its vertical position using a metal rod driven by a shaker. The rod produces a single impact at the floater center. As sketched in Fig. \ref{fig:Schematic_resonance_frequency_setup}a), the impact triggers vertical oscillations. 
\begin{figure}[h!]
    \centering
    \includegraphics[width=0.99\columnwidth]{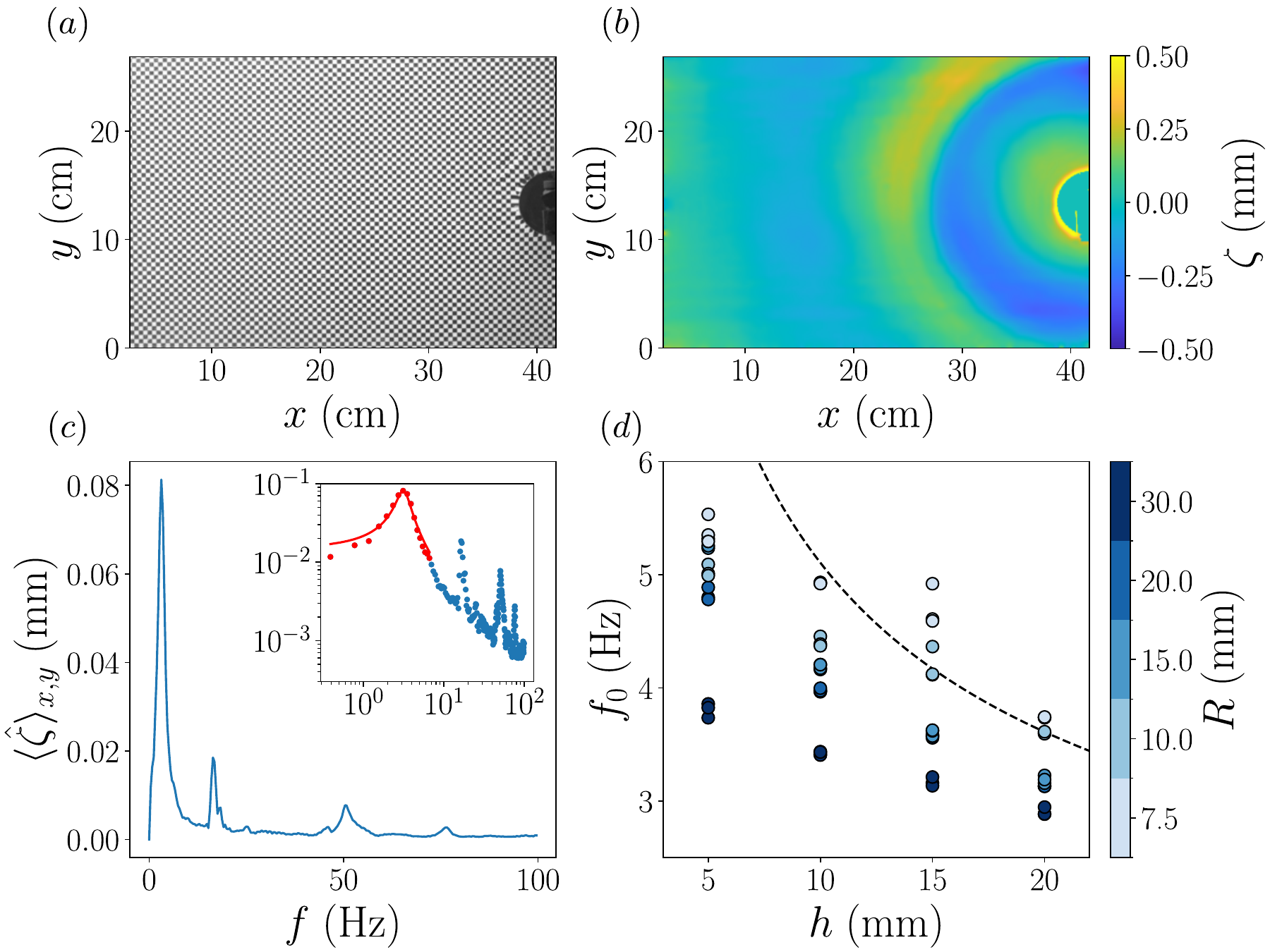}
    \caption{Wave field and resonance frequency of floating cylinders experiencing instantaneous vertical impacts. (a) Top view of a floating cylinder ($R = 30 \, \rm mm$, $h = 15 \, \rm mm$) displaced by a rod (as depicted in the setup of Fig. \ref{fig:Schematic_resonance_frequency_setup}a) over a checkerboard pattern with square size $a_{c} = 4.0 \, \rm mm$. (b) Water surface topography at time $t = 1.39 \, \rm s$ post impact. (c) Fourier spectrum $|\langle \hat{\zeta} \rangle_{x,y}| (f)$ of the wave field generated by the object perturbation shown in (b). \correctref{The inset in (c) depicts the same Fourier spectrum in log-log scale. The main peak represented by red dots is fitted with a Lorentzian function, plotted as a red line.} (d) Measured resonance frequency as a function of the floaters' height $h$ and for different radii $R$. Each data point corresponds to an individual experiment. \correctref{The black dotted line represents the prediction $f_0 = \frac{1}{2\pi}\sqrt{\frac{\rho_w g}{\rho h}}$ obtained without considering the floater added mass.}}
    

    \label{fig:Height_field_and_TF_spectrum}
\end{figure}

To prevent the floater from drifting, we loosely attach its upper edge to the tank walls using two nylon wires without affecting its vertical motion. The wave topography generated by the floater oscillations is measured using a technique named Fast Checkerboard Demodulation (FCD)~\cite{wildeman_real-time_2018}. A checkerboard pattern is placed beneath the tank, and a camera records 200 frames per second on the surface from above. We then measure the pattern distortion for each video frame and compare it to a static reference image. The checkerboard pattern distortion is caused by light refraction at the air-water interface and a reconstruction of surface elevation with a sensitivity of $\approx 10 \, \mu\text{m}$ (see Fig \ref{fig:Height_field_and_TF_spectrum}a) can be achieved. Wave propagation occurs in the deep water regime ($kH\gg1$, with $k$ the wavenumber). The checkerboard square size $a_{c} = 4.0 \,\rm{mm}$ is selected to avoid surface reconstruction errors following the criteria of Moisy {\textit{et al.}} \cite{moisy_synthetic_2009}. 

The wavefield is recorded from the instant of impact until the waves reflect from the tank boundaries. The reconstructed surface $\zeta(x, y, t)$ (see Fig. \ref{fig:Height_field_and_TF_spectrum}b) is analyzed using Fourier analysis. We first compute $\hat{\zeta}_{x,y} (f)$, the Fourier spectrum in time at all (spatial) locations of the 2D reconstructed wave field. We then average the Fourier spectra in space over the entire reconstructed area. Fig. \ref{fig:Height_field_and_TF_spectrum}c shows the absolute value of $< \hat{\zeta}_{x,y} (f) >$ as a function of frequency $f$ in logarithmic scale. For all our experiments, we observe one prominent peak with frequency in the range $2 - 6$~Hz and a few secondary peaks at much larger frequencies (typically for  $f>15$ Hz). 

\correctref{This prominent peak is normalized and approached by a Lorentzian function $\frac{1}{\sqrt{1 + 4\pi^2 \left(\frac{f - f_0}{\mu} \right)^2}}$. From this fit, we extract with good precision the frequency of maximum emission $f_0$ as well as the damping coefficient $\mu$. This coefficient accounts for several dissipative phenomenons such as bulk and boundary viscous effects, capillary effects at the object walls and waves radiation. Values of this damping coefficient range between $2$ and $4 \; \mathrm{rad.s^{-1}}$ resulting in damped oscillations. However, the frequency shift induced by this energy loss is small, resulting in a relative shift in the order of $2 \; \%$ on the natural resonance frequency. We therefore consider that the frequency of maximum emission $f_0$ we measured is a good proxy of the natural resonance frequency of the floater.}

Figure \ref{fig:Height_field_and_TF_spectrum}d shows the frequency $f_0$ of maximum emission as a function of the cylinder thickness $h$ and for different radii. \correctref{Each point corresponds to a single experiment. For a single cylinder geometry, three different experiments are performed, resulting in a mean standard deviation of about $0.04 \; \mathrm{Hz}$}. We observe that $f_0$ decreases with both the cylinder thickness and the cylinder radius.
This indicates that both parameters play a role in determining the heaving resonance frequency $f_0$.

\subsection{Modeling}\label{sec:modeling}

Historically, floating objects have been studied, most notably in naval contexts, where engineers sought to understand the motion of ships. Complex variable techniques have been exploited to calculate the added mass of two-dimensional hulls atop a fluid of infinite depth as early as 1929 \cite{lewis1929inertia}. An improved investigation of the heaving motion of two-dimensional bodies was performed in later studies \cite{ursell_heaving_1949,yu1961surface,bai_added_1977}, which also consider finite-depth effects. The motion of three-dimensional floating bodies has also been investigated in the case of vertical circular cylinders \cite{yeung_added_1981} and spheres \cite{havelock_waves_1955}. The referenced works rely mainly on expansion formulae and integral expressions in order to solve for the ensuing motion of floating obstacles. 

\begin{figure}[h!]
    \centering
    \includegraphics[width = 0.7\textwidth]{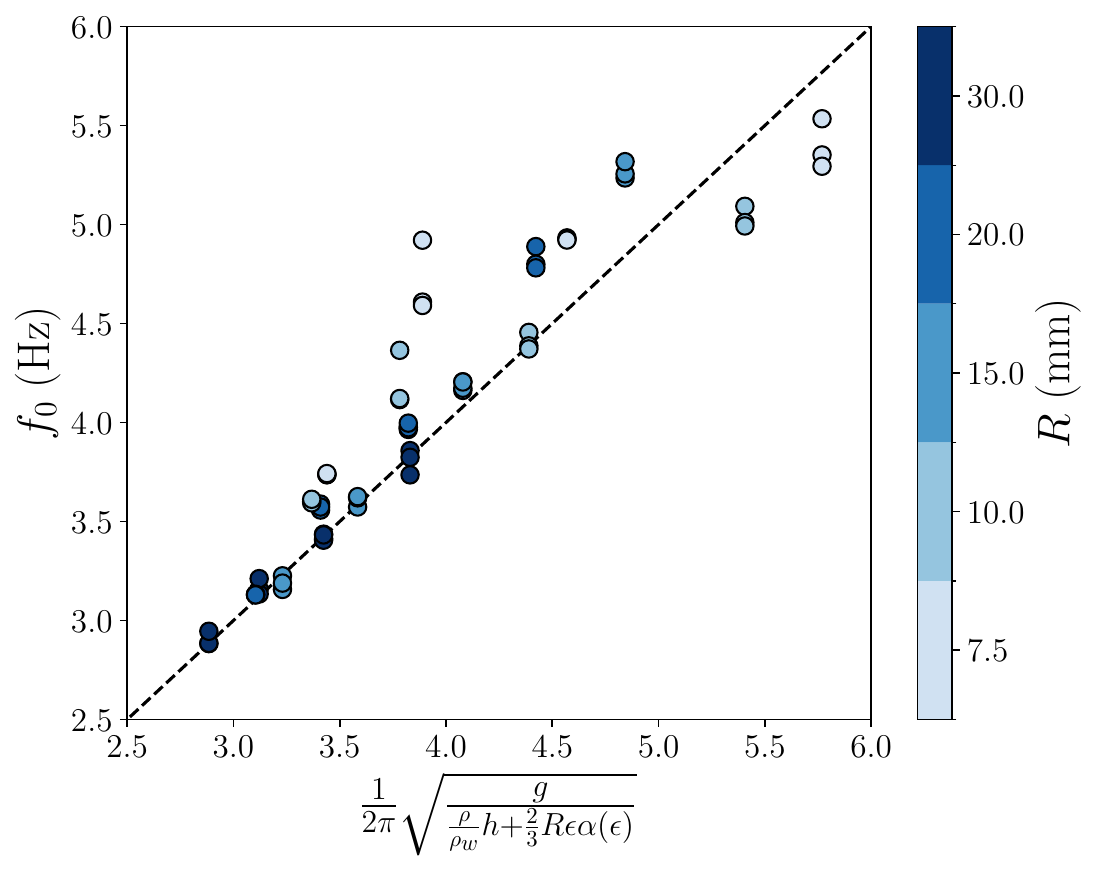}
    \caption{Experimental data compared to our theoretical model, using (\ref{eq:true}). Each point in the plot represents a single experiment. Cylinder radii are given by color code. For a given radius, cylinder heights are, from right to left in the plot: $h = 5 \, \rm{mm}$, $h = 10 \, \rm{mm}$, $h = 15 \, \rm{mm}$, $h = 20 \, \rm{mm}$.}
    \label{fig:Data_comparison_epsilon}
\end{figure}

\correctref{
In the present work, we utilize similar theoretical notions. In the limit of high-frequency oscillations, and using a simplifying geometric approximation, we derive an expression for the added mass and deduce the natural resonance frequency of oscillation of a cylinder. Theses approximations agree well with our data (see Fig. \ref{fig:Data_comparison_epsilon}).
}

\begin{figure}[t!]
    \centering
    \includegraphics[width =0.99\textwidth]{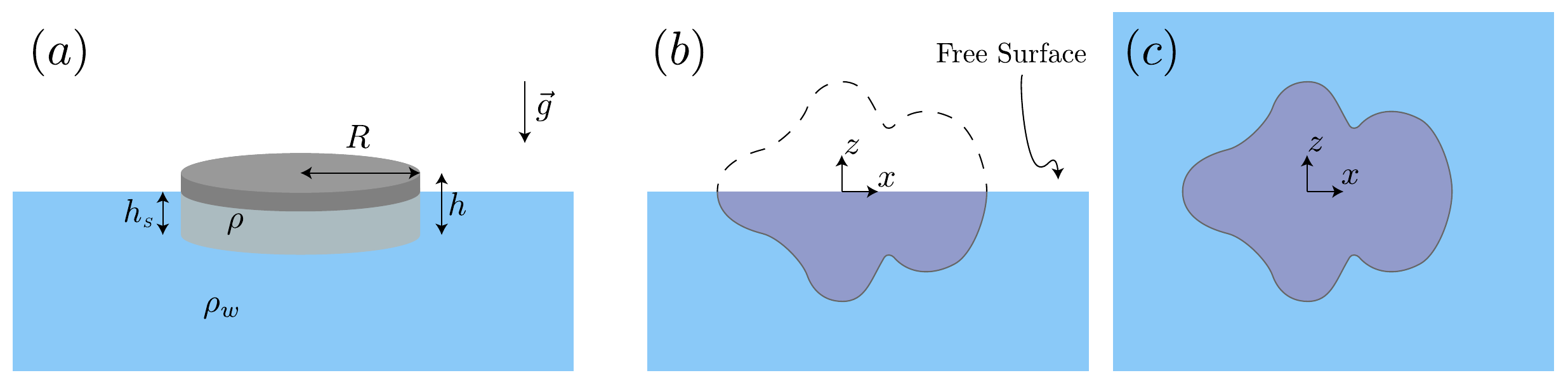}
    \caption{Schematics for the model. (a) Floating cylinder. In the high-frequency limit, the heaving added mass of the shape in panel (b) is equal to half the translatory added mass, $M_{22}$, of the shape in panel (c) in an infinite fluid. The shape in (c) is the union of the submerged portion of the body in (b) and its reflection about the free surface.}
    \label{fig:Schematic_floating_cylinder}
\end{figure}

We describe heave oscillations of a buoyant cylinder floating on the free surface of a water bath with density $\rho_w$. 
We denote the cylinder's submerged height with $h_s$. (Fig. \ref{fig:Schematic_floating_cylinder}a). 

\correctref{We consider a vertical unit vector $\vec{e_z}$ oriented upward. At rest, the cylinder's weight $-\rho \pi R^2 h g \vec{e_z}$ and the resultant of buoyancy forces $\rho_w \pi R^2 h_s g\vec{e_z}$ compensate, resulting in the expression of the submerged depth $h_s = \frac{\rho}{\rho_w}h$} \\


For a small perturbation $\delta z(t)$ from the equilibrium position, a naive force balance on the floating object gives $\rho \pi R^2 h \partial_{tt} {\delta z}=-\rho_w \pi R^2 g \delta z$, which yields the natural frequency of oscillation $\Omega=\sqrt{\rho_w g / (\rho h)}$. However, such a calculation overlooks two important considerations. First, the surrounding fluid must be accelerated along with the cylinder, increasing the effective inertia. Second, the waves at the free surface impose a reaction force on the cylinder. \\

\textbf{The High Frequency Limit}\\

The boundary value problem for a heaving cylinder is inherently complex, even when assuming potential flow and linearized boundary conditions. In the following, we use a high-frequency approximation to derive an asymptotic expression for the resonant frequency. We take the limit $\Omega ^2 L/g\rightarrow \infty$, with $L$ the typical length of the floating body, $g$ the acceleration due to gravity, and $\Omega$ the oscillation frequency. The problem then simplifies as the boundary condition on the free surface reduces to $\phi = 0$, where $\phi$ is the velocity potential, while the standard boundary conditions remain applicable elsewhere (see \citet{bai_added_1977} and \citet[\S 2]{ursell1953short}). Note that for a typical experiment, with $L=2R $ and $\Omega=2\pi f_0 $, giving $\Omega^2 L /g \sim 1-4$. Though these values are not extremely large, we still take the high-frequency limit as the first approximation to our problem. 

We use the high frequency limit to compute the added mass by reflecting the submerged portion of a body and then halving the resultant added mass, referred to as the method of duplicated models \cite[p. 135]{korotkin2008added} (see the sketch in figure \ref{fig:Schematic_floating_cylinder}(c)). For a vertical translation of the body at some velocity, $\phi(x,y,-z)=-\phi(x,y,z)$ by symmetry, which implies $\phi(x,y,0)=0$. As a consequence, the potential $\phi$ around the shape sketched in figure \ref{fig:Schematic_floating_cylinder}(c) restricted to the domain $z<0$ is a solution to the boundary value problem outlined by Bai \cite{bai_added_1977}. Since the flow magnitude is symmetric about $z=0$, half the energy of the flow around the translating shape in figure \ref{fig:Schematic_floating_cylinder}(c) is in the upper half plane and half is in the lower half-plane. Thus, the added mass for the body in figure \ref{fig:Schematic_floating_cylinder}(b) is precisely half of the added mass for the shape in figure \ref{fig:Schematic_floating_cylinder}(c). 
\\

\textbf{Added Mass of Submerged Cylinders: An Ellipsoidal Approximation }\\

To compute the added mass of a floating cylinder of radius $R$ and submerged portion $h_s$, we instead compute the added mass of an immersed cylinder of height $2 h_s$ and then divide by a factor of two, according to the method presented in the previous section. To avoid complications arising at the cylinder corner, we approximate the cylinder by an ellipsoid with semi-axis $h_s$ along the vertical direction and radii $R$ along the other two directions. We compute the potential flow around an ellipsoid traveling along its axis at speed $U$ as the solution to a Laplace problem in ellipsoidal coordinates. Next, the kinetic energy of the flow is computed explicitly \correctref{(see Lamb \citep{lamb1945hydrodynamics}[Eq. 11 of \S 114])}. By considering the pure translational motion of the ellipsoid along its semi-axis of radius $h_s$, the translational added mass $M_{22}$ becomes,
\begin{equation}
    M_{22}=\rho_w\frac{4 \pi h_s R^2}{3}\frac{\sqrt{1/\epsilon^2-1}-\acos{\left(\epsilon\right)}}{\acos{\left(\epsilon\right)-\epsilon \sqrt{1-\epsilon^2}}},
\end{equation}
where $\epsilon \equiv h_s/R$.
Considering the additional movement of the surrounding fluid, we now perform the force balance, including the effect of the added mass. The equation of motion then becomes,

\correctref{
\begin{equation}
    (\rho \pi h R^2  +\frac{1}{2}M_{22})\ddot{\delta z} + \rho_w \pi R^2 g \delta z = 0.
\end{equation}
}

From this equation, we deduce the natural frequency of oscillation of the body, 
\begin{equation}\label{eq:true}
\Omega = \sqrt{\frac{ \rho_w g \pi R^2 }{\rho h \pi R^2 + M_{22}/2}}=\sqrt{\frac{  g  }{r h  + \frac{2 R}{3}\epsilon \alpha{\left(\epsilon \right)}}},
\end{equation}
where $r=\rho/\rho_w$ and
\begin{equation}
\alpha{\left(\epsilon\right)}=\frac{\sqrt{1/\epsilon^2-1}-\acos{\left(\epsilon\right)}}{\acos{\left(\epsilon\right)-\epsilon \sqrt{1-\epsilon^2}}}.
\end{equation}
Note that, in this formulation, we have neglected the memory convolution term that becomes important as the wave field builds up \cite[see Eq. 31]{newman1985transient}, and thus the frequency is expected to be valid on short time scales or in situations in which the history integral remains small.

We compare the theoretical prediction with the cylinder resonance frequency $f_0$ measured experimentally in Fig \ref {fig:Data_comparison_epsilon}. Our model is in good agreement with the experimental data. The agreement is robust for smaller values of $\Omega$, 
\correctref{obtained for the larger tested objects. We observe that the experimental results deviate from the theory for smaller radii, typically for a radius of R = 7.5 mm. This deviation may correspond to the fact that for elongated vertical cylinders, the assumption of pure heaving motion may break down.}
\section{FLOATING OBJECT IN AN EXTERNAL WAVEFIELD} 

\begin{figure}[h!]
    \centering
    \includegraphics[width=\columnwidth]{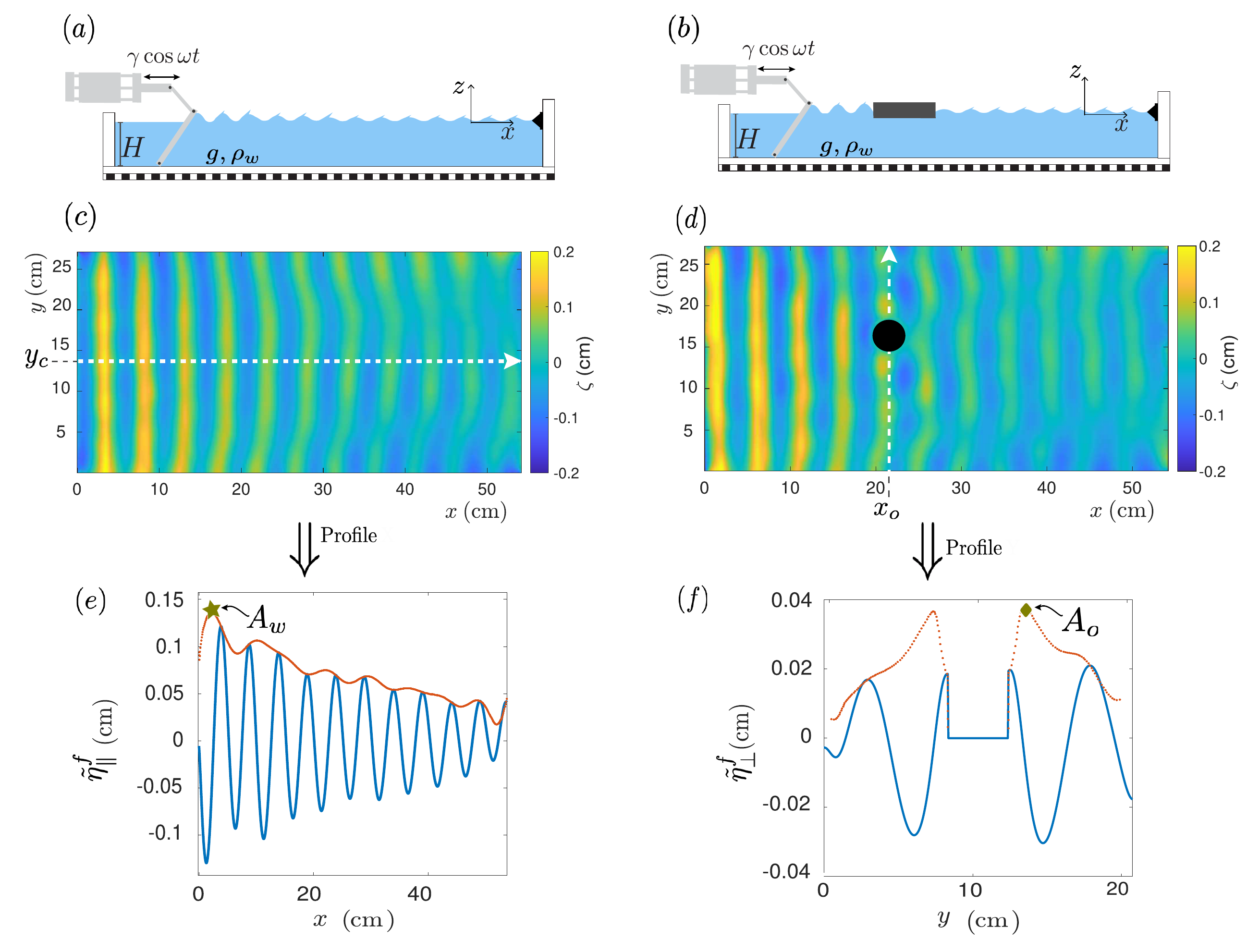}
    \caption{Experimental setups and sample data for characterizing the wave field in the vicinity of a floating object: configuration without (a,c,e) and with (b,d,f) cylindrical floater. (a, b) Schematics of the experimental setups (not to scale), in which a wavemaker generates harmonic waves in the tank. (c, d) Sample instantaneous height fields $ \zeta(x,y) $ with wavemaker frequency $f = 6.0 \, \rm Hz$. In (d) the floating cylinder has radius $R = 20 \, \rm mm$ and height $h = 10 \, \rm mm$. (e) Processed spatio-temporal profile along the dashed line at $y=y_c$ in (c). (f) Processed spatio-temporal profile along the dashed line at $x=x_o$ in (d), passing through the center of the floating object. Blue lines show the wave height at a given time, while red lines are the envelopes of the processed signal.}
    \label{fig:Schematics_extraction_profile}
\end{figure}
   
\begin{figure}[h!]
    \centering
    \includegraphics[width=\columnwidth]{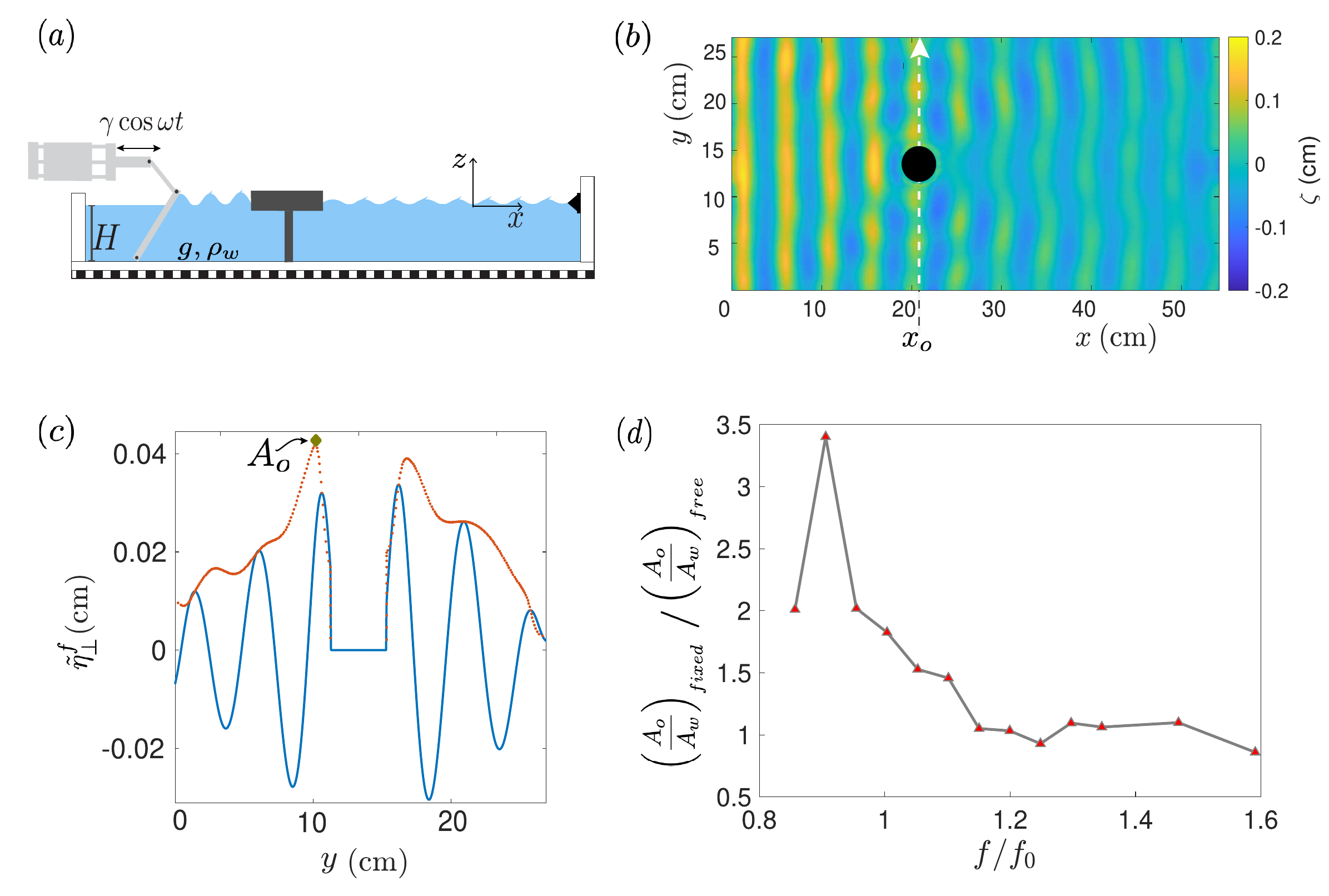}
    \caption{Experimental setup and sample data for characterizing the wave field in the vicinity of a fixed object. (a) Schematic of the experimental setup (not to scale). The floater is attached to the water tank via a metal stick. (b) Snapshot of the reconstructed wave field with forcing frequency $f = 6.0 \, \rm Hz$. The floater has radius $R = 20 \, \rm mm$ and height $h = 10 \, \rm mm $. (c) A filtered spatio-temporal profile along the dashed line at $x=x_o$ in (b), passing through the center of the object. The blue line shows the wave height at a given time, while the red line represents the processed signal's envelope. The maximum of the envelope is noted $A_o$. (d) Ratio of the normalized amplitudes of wave emitted in the transverse direction by the fixed and the free object as a function of the normalized forcing frequency $f/f_0$. 
    }
    \label{fig:Free_objects_VS_clamped}
\end{figure}

We now examine the interaction between the floater and externally generated surface waves. The experiment is designed to measure the wave field near the floating cylinder. We compare two configurations: in the first configuration, the object can move freely at the surface in response to the incoming wave field (see Fig. \ref{fig:Schematics_extraction_profile}b); in the second, the object is attached to the bottom of the water tank with a rigid stick (Fig. \ref{fig:Free_objects_VS_clamped}a). 
We generate the surface waves using a plate connected to a linear motor, as sketched in Fig \ref{fig:Schematics_extraction_profile}. A wave absorber, made of triangular shapes of expanded polystyrene, is placed at the perimeter of the interface to reduce wave reflection. The wave field is recorded at 100 frames per second for 10 seconds. The wave field $\zeta(x, y, t)$ is measured with the surface reconstruction technique described in Section II~\cite{wildeman_real-time_2018}.

In the free-floating configuration, a cylinder is placed 10 cm away from the wave generator and farther from the lateral edges. We systematically vary the wave frequency $f$ while maintaining a constant plate displacement amplitude. Figs. \ref{fig:Schematics_extraction_profile}c and \ref{fig:Schematics_extraction_profile}d show the comparison between the wave field with and without the object. In the presence of the object, we observe the superposition of the incoming wave, the reflected waves, and the waves emitted by the oscillations of the cylinder. We focus on wave emission near the natural frequency $f_0$ of the floater.
\begin{figure}[h!]
    \centering
    \includegraphics[width=\columnwidth]{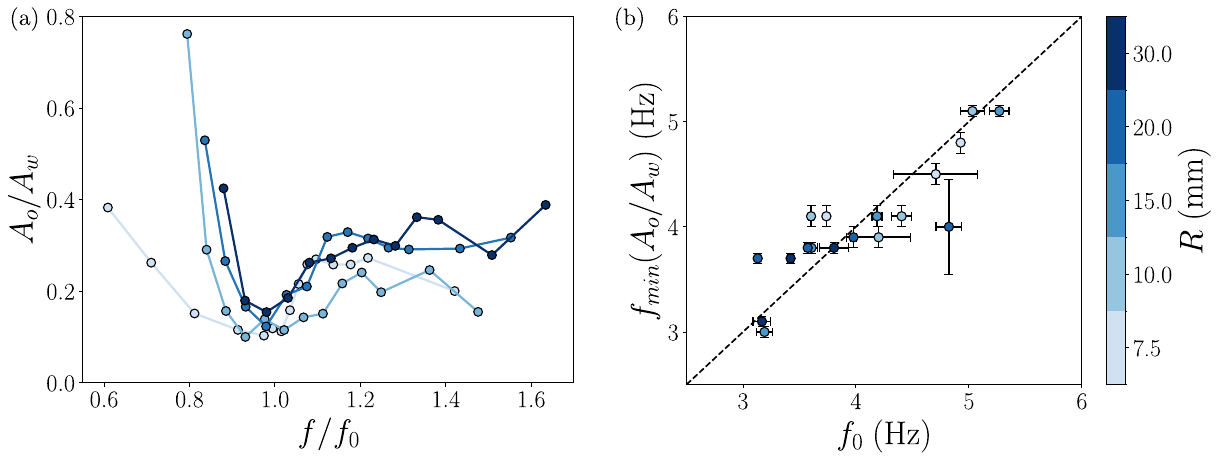}
    \caption{Effect of the object on the incoming wave field. (a) Normalized amplitude $A_o/A_w$ of the waves emitted in a direction orthogonal to the incoming wave vector as a function of the normalized frequency of the incoming waves, $f/f_0$, for various radii $R$ and fixed height $h=10 \: \rm mm$. (b) Comparison between the frequency at which the normalized amplitude is minimal, $f_{\rm min}(A_o/A_w)$, and the natural frequency $f_0$ of the cylinders for all the tested geometries. }
\label{fig:Reemitted_amplitude_ratio_and_frequencies_correspondence}
\end{figure}
First, we extract from the wave field a one-dimensional spatiotemporal profile in the absence of the floater, ${\eta}_\parallel(x,t) = \zeta(x, y=y_c, t)$, parallel to the generated wave vector, where $y=y_c$ is the tank centerline (Fig.\ref{fig:Schematics_extraction_profile}c). After filtering in time and space, we obtain $\tilde{\eta}_\parallel^f(x,t)$ (refer to supplementary material, section II A, for details on signal processing). Computing the envelope of $\tilde{\eta}_\parallel^f(x,t)$ yields the maximum wave amplitude $A_w(\tilde{\eta}_\parallel^f)$, at each frequency (see Fig. \ref{fig:Schematics_extraction_profile}e). This serves as a reference for later comparisons with the wave field in the presence of the object. We proceed by analyzing the interaction between the incoming wave field and the free-floating cylinder. We extract spatio-temporal profiles ${\eta}_\perp(y,t) = \zeta(x=x_o, y, t)$, perpendicular to the generated wave vector, where $x=x_o$ is the coordinate of the object's center of mass (Fig.\ref{fig:Schematics_extraction_profile}d). After filtering, we obtain $\tilde{\eta}_\perp^f(y,t)$, which represents only the waves re-emitted by the object in response to the incoming wave field, as isolated through the filtering process (refer to supplementary material, section II B, for details on signal processing). From this filtered signal, we extract the maximum wave amplitude  $A_o(\tilde{\eta}_\perp^f)$ (Fig. \ref{fig:Schematics_extraction_profile}f). 

The ratio of re-emitted to undisturbed wave amplitudes, $A_o/A_w$, is used to quantify the object's influence on the incoming wave field. 
The ratio $A_o/A_w$ is measured for multiple object geometries varying both the radius $R$ and the height $h$ of the cylinder. Fig. \ref{fig:Reemitted_amplitude_ratio_and_frequencies_correspondence}a shows the amplitude ratio as a function of the excitation frequency $f/f_0$. We observe that the wave emission in the transverse direction is minimal close to the floater's resonance frequency $f_0$. This behavior suggests that the object absorbs the wave energy at its natural frequency, decreasing the wave emission in the transverse direction. A similar trend is observed for a fixed height (see supplementary material, section III). 
For each floater, we extract the frequency $f_{\rm min}$ corresponding to the minimum of emission in the transverse direction. Figure \ref{fig:Reemitted_amplitude_ratio_and_frequencies_correspondence}b shows $f_{\rm min}$ as a function of the natural frequency $f_0$ of the floater. We observe that $f_0$ is a proxy for $f_{\rm min}$ for all the geometries considered.

To further highlight the effect of the object motion on the wave emission, we conducted experiments in the second configuration, that is, with a cylinder fixed to the bottom of the tank (Fig. \ref{fig:Free_objects_VS_clamped}a,b,c). The cylinder has the same geometry as the free-floating cylinders. The ratios $A_o/A_w$ of free and fixed cylinders are compared in Fig. \ref{fig:Free_objects_VS_clamped}d. Near resonance, the fixed object exhibited greater transversal wave emission than the free-floating object. The two configurations show a similar response for an excitation frequency higher than $1.2f_0$. These observations support the hypothesis that at resonance, the free floater absorbs and redistributes wave energy unevenly. The floater radiates energy, leading to interactions between the emitted and diffracted wave fields, which may explain the reduced emission efficiency. In contrast, the fixed object, unable to oscillate, only reflects and diffracts the incoming waves.\\

\section{CONCLUSION} 
We have investigated the interaction of a floating cylinder with surface waves in laboratory-scale experiments. In the first part, the bodies' natural oscillation frequencies were experimentally characterized \correctref{thanks to a demodulation method which enabled the reconstruction of the surface deformation with a high resolution in both space and time.} We show that the resonance frequency of the heaving motion corresponds to a maximum wave emission. The resonance frequency $f_0$ depends on the cylinder radius $R$ and its thickness $h$. We present a theoretical model that rationalizes the observed resonance frequencies for all the body dimensions characterized experimentally. We show that, in the high-frequency limit, the added mass of a floating body corresponds to half the same body's added mass when fully submerged in the fluid bulk. \correctref{Although the high-frequency hypothesis is not strictly fulfilled for the heaving motions in our experiments, the comparison with the experimental resonant frequency is quantitative.} 
In the second part, we studied the effects of the floater's vertical motion on the re-emitted wave field. We measured the wave field re-emitted by the floating cylinder and showed that, in the transverse direction, the re-emitted wave exhibits a minimum emission for an excitation frequency close to the body resonance frequency $f_0$. By comparing the wave fields re-emitted by a fixed and a free-floating cylinder with the same dimension, we proved that the body oscillations are indeed responsible for the observed minimal transverse emission.
\\

\begin{acknowledgements}
This work has benefited from the financial support of Mairie de Paris through Emergence(s) grant 2021-DAE-100 245973, and the Agence Nationale de la Recherche through grant MSIM ANR-23-CE01-0020-02. W.R. is supported by the Italian Ministero dell'Universit\`a e della Ricerca (MUR) through the program PON Ricerca e Innovazione 2014-2020.
\end{acknowledgements}

\begin{figure}[h!]
    \centering
    \includegraphics[width = 0.7\textwidth]{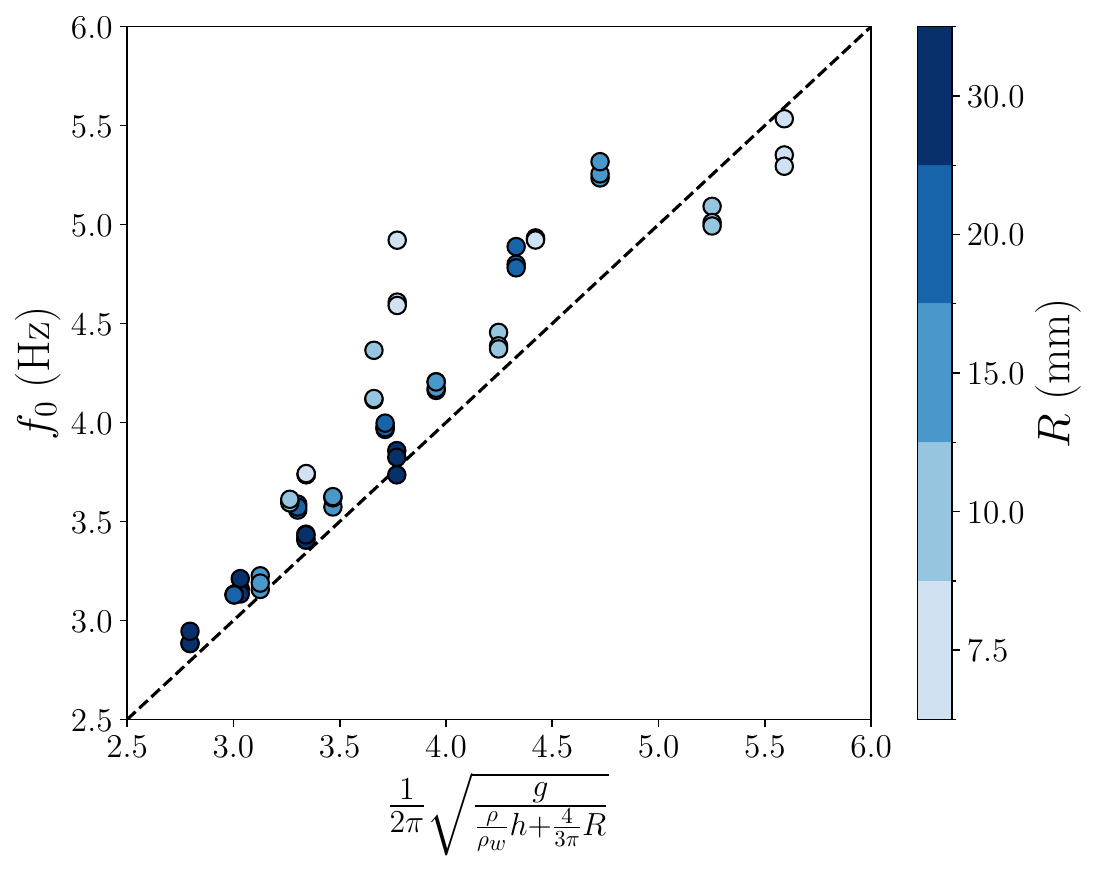}
    \caption{Experimental data compared to our theoretical model, using the approximated formula (\ref{eq:aprox}). Each point in the plot represents a single experiment.Cylinder radii follow the color bar on the right. For a given radius and from right to left in the plot, cylinder heights are: $h = 5 \, \rm{mm}$, $h = 10 \, \rm{mm}$, $h = 15 \, \rm{mm}$, $h = 20 \, \rm{mm}$.}
    \label{fig:Data_comparison}
\end{figure}

\newpage
\maketitle
\begin{center}
\textbf{SUPPLEMENTAL MATERIAL}
\end{center}
~\\
\begin{center}
\textbf{I. The Disk Limit: A Uniform Approximation}
\end{center}
To study the case of a thin floating disk, we can take the limit $\epsilon\rightarrow 0$. In the expression of the resonant frequency of the main text, we see that 
the product $\epsilon\alpha(\epsilon)$ tends to $2/\pi$. Thus, for small values of $\epsilon$, the frequency reduces to
\begin{equation}\label{eq:aprox}
\Omega(\epsilon \rightarrow 0) \rightarrow \sqrt{\frac{  g  }{r h  + \frac{4 R}{3\pi}}}.
\end{equation}
Naively using this formula over the whole range of experiments, even though in many cases $\epsilon = \mathcal{O}\left(1\right)$, we find a good agreement with the experiment, as depicted in figure \ref{fig:Data_comparison}.

One might then ask the following question: \emph{for arbitrary $\rho,\rho_w, R,h$, how well does $\Omega(\epsilon = 0)$ approximates the natural frequency more generally when $\epsilon \neq 0$?} To answer this question, we examine the ratio $\Omega(\epsilon=0)/\Omega(\epsilon)$. When this ratio is equal to unity, the approximation is exact. We compute
\begin{equation}
\frac{\Omega(\epsilon=0)}{\Omega(\epsilon)}=\sqrt{\frac{r h+\frac{2R}{3}\epsilon \alpha\left(\epsilon\right)}{r h +\frac{4R}{3\pi}}}=\sqrt{\frac{1+\frac{2}{3} \alpha\left(\epsilon\right)}{1+\frac{4}{3\pi\epsilon}}},
\end{equation}
which only depends on $\epsilon$. Furthermore, by plotting this ratio as a function of epsilon, we see that 
$$0.96<\Omega(\epsilon=0)/\Omega(\epsilon)<1$$
for all values of $\epsilon$. Remarkably, one may invoke the approximation outlined in (\ref{eq:aprox}) for any density ratio $r$, radius $R$, and height $h_0$ and be guaranteed that the error, as compared to the solution of equation (\ref{eq:true}), is less than about 4\%. Thus, equation (\ref{eq:aprox}) represents a very valuable approximation for estimating the natural vibration of a floating cylinder.

~\\
\begin{center}
\textbf{II. Signal processing for profile extraction}
\end{center}
\begin{center}
\textbf{II A. Parallel profile}
\end{center}

The spatiotemporal profile reported in section III of the main text is processed in the Fourier space \cite{gonzalez2002processing} to filter the waves reflected by the tank's boundaries and mitigate spurious fluctuations associated with low wave numbers. This Fourier filtering process involved transforming the profiles \( \eta_{\parallel}(x,t) \) into the spatial and temporal frequencies domain using the Fourier transform \( \mathcal{F} \); applying a filter function \( H(k,\omega) \) to suppress undesired frequencies; transforming the filtered profiles back to the spatio-temporal domain using the inverse Fourier transform \( \mathcal{F}^{-1} \). These steps can be written as:
\[ \hat{\eta}_{\parallel}(k,\omega) = \mathcal{F}[\eta_{\parallel}(x,t)], \]
\[ \tilde{\eta}_{\parallel}(x,t) = \mathcal{F}^{-1}[H(k,\omega)\hat{\eta}_{\parallel}(k,\omega)], \]
where $\hat{\eta}_{\parallel}(k,\omega)$ corresponds to the Fourier transform of the horizontal profile and $\tilde{\eta}_{\parallel}(x,t)$ represents the filtered spatio-temporal signal. This filtered signal is then demodulated at the excitation frequency $f$, leading to the function $\tilde{\eta}_{\parallel}^f(x,t)$. A frame of this demodulated profile is represented as the blue curve in figure \ref{fig:Schematics_extraction_profile}e in the main text. To measure the incoming wave amplitude, we extract the envelope of this demodulated profile via a peak detection algorithm that detects the local maxima of the demodulated curve $\tilde{\eta}_{\parallel}^f(x,t)$ for all time steps. The envelope curve $E_\parallel(\tilde{\eta}_{\parallel}^f)$ is represented in red in figure \ref{fig:Schematics_extraction_profile}e in the main text. 

\begin{center}
\textbf{II B. Perpendicular profile}
\end{center}First, a filter $H_{r}(k,\omega)$ is applied to the transverse profile in the Fourier space to filter the waves reflected by the boundaries of the tank. Then, a Butter-worth filter function $H_{bw}(k,\omega)$ is applied to suppress undesired frequencies at low and high wave numbers \cite{damiano2016surface,surferbot}. This filtered signal is then transformed back to a spatiotemporal signal $\tilde{\eta}_{\perp}(y,t)$ using the inverse Fourier transform $\mathcal{F}^{-1}$: 
\[ \hat{\eta}_{\perp}(k,\omega) = \mathcal{F}[\eta_{\perp}(y,t)], \]
\[ \tilde{\eta}_{\perp}(y,t) = \mathcal{F}^{-1}[H_{r}(k,\omega)H_{bw}(k,\omega)\hat{\eta}_{\perp}(k,\omega)]. \]
Finally, the profile $\tilde{\eta}_{\perp}(y,t)$ is demodulated at the excitation frequency $f$ to obtain the spatio-temporal signal $\tilde{\eta}_{\perp}^f(y,t)$, plotted as the blue curve in figure \ref{fig:Schematics_extraction_profile}f in the main text. The maximum amplitude of the waves emitted by the object in this direction is obtained by extracting the envelope of $\tilde{\eta}_{\perp}^f(y,t)$, as described in the previous section. The envelope $E_\perp(\tilde{\eta}_{\perp}^f)$ is represented in red in figure \ref{fig:Schematics_extraction_profile}f in the main text.\\\\

\begin{center}
\textbf{ III. Effect on wave field from cylinders with fixed height}
\end{center}

\begin{figure}[h!]
    \centering
    \includegraphics[width = 0.7\textwidth]{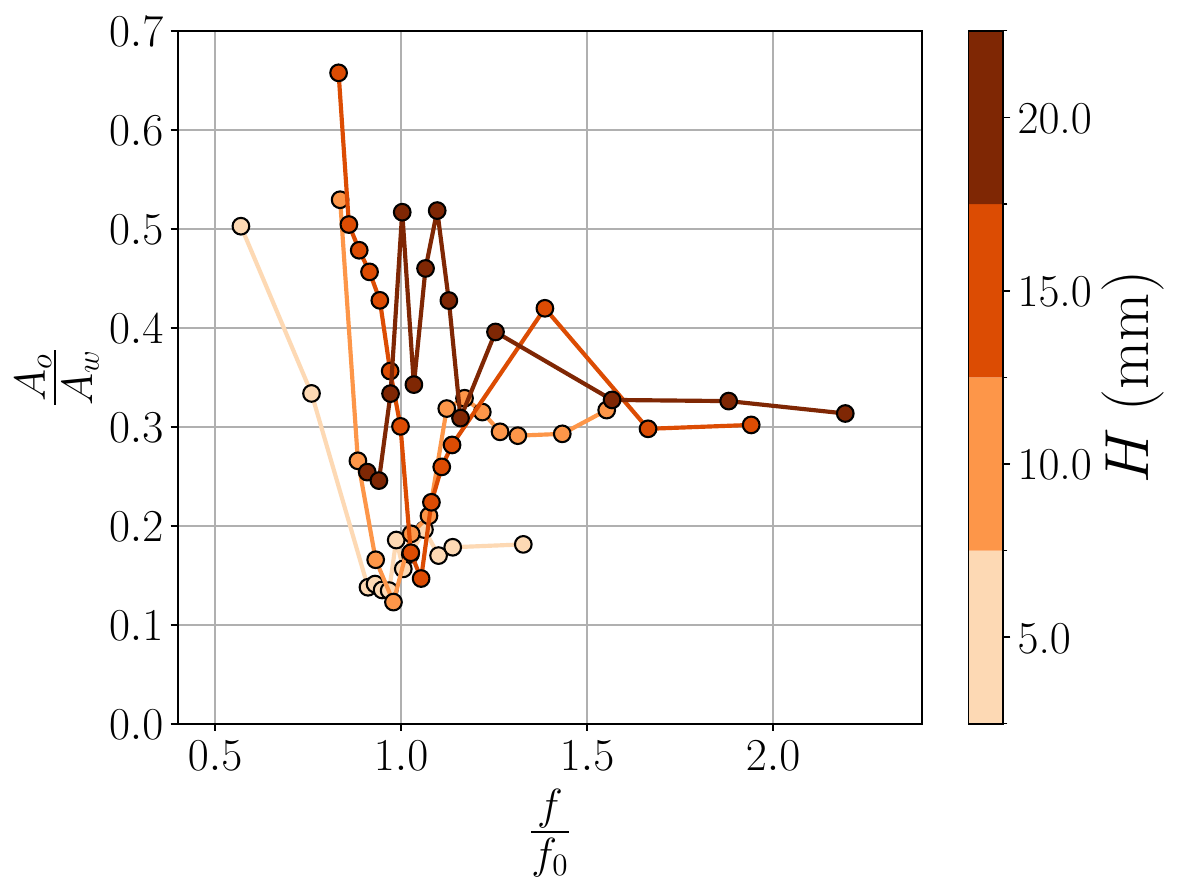}
    \caption{Normalized amplitude $A_o/A_w$ of the waves emitted in a direction orthogonal to the incoming wave vector as a function of the normalized frequency of the incoming waves, $f/f_0$, for various heights $h$ and fixed radius $R=15 \: \rm mm$.} 
    \label{fig:Supp_Matrial_fixedradius}
\end{figure}

Using the same approach as in Figure \ref{fig:Reemitted_amplitude_ratio_and_frequencies_correspondence} of the main text, the ratio $A_o/A_w$ was determined for different heights of the free-floating cylinders with fixed radius $R = 15 \: \rm mm$. The results are plotted in Figure \ref{fig:Supp_Matrial_fixedradius} as a function of the excitation frequency $f$ normalized by the measured resonance frequency $f_0$ of each body. Here, the system also exhibits a minimum value of the ratio $A_o/A_w$ at an excitation frequency close to the resonance frequency. Therefore, near its resonance frequency $f_0$, a cylindrical object shows a reduced emission efficiency in the direction orthogonal to the initial wave propagation.

\end{document}